\documentclass[english,prl,reprint,nofootinbib,longbibliography,notitlepage,superscriptaddress,preprintnumbers,twocolumn,showkeys]{revtex4-1}
\usepackage[latin9]{inputenc}
\usepackage{graphicx}
\usepackage{hyperref}
\usepackage{amsmath}
\usepackage{amssymb}
\usepackage{microtype}
\usepackage{xcolor}
\usepackage{cleveref}
\usepackage{bbm}
\usepackage{mathtools}
\usepackage{todonotes}

\newcommand{\Mp}{M_\mathrm{Pl}}

\newcommand{\Meff}{M_\mathrm{eff}}

\newcommand{\para}[1]{\par\vspace{2mm}\noindent{\bf{{#1}}}\,---\,}

\makeatletter
\DeclareRobustCommand{\rcite}[1]{%
  \rcite@aux#1,\@nil{#1}%
}
\def\rcite@aux#1,#2\@nil#3{%
  \if\relax#2\relax
    Ref.~\cite{#3}%
  \else
    Refs.~\cite{#3}%
  \fi
}
\makeatother

\hypersetup{
    colorlinks = true,
    citecolor = {red},
    linkcolor = {blue},
    urlcolor = {blue},
}

\begin{document}

\title{Multi-field dark energy: cosmic acceleration on a steep potential}

\author{Yashar Akrami}
\email{akrami@ens.fr}
\affiliation{Laboratoire de Physique de l'\'Ecole Normale Sup\'erieure, ENS, Universit\'e PSL, CNRS, Sorbonne Universit\'e, Universit\'e de Paris, F-75005 Paris, France}
\affiliation{Observatoire de Paris, Universit\'e PSL, Sorbonne Universit\'e, LERMA, 75014 Paris, France}

\author{Misao Sasaki}
\email{misao.sasaki@ipmu.jp}
\affiliation{Kavli Institute for the Physics and Mathematics of the Universe (WPI), UTIAS, The University of Tokyo, Chiba 277-8583, Japan}
\affiliation{CGP, Yukawa Institute for Theoretical Physics, Kyoto University, Kyoto
606-8502, Japan}
\affiliation{LeCosPA, National Taiwan University, Taipei 10617, Taiwan}

\author{Adam R. Solomon}
\email{adamsolo@andrew.cmu.edu}
\affiliation{Department of Physics \& McWilliams Center for Cosmology,\\Carnegie Mellon University, Pittsburgh, Pennsylvania 15213, USA}

\author{Valeri Vardanyan}
\email{valeri.vardanyan@ipmu.jp}
\affiliation{Kavli Institute for the Physics and Mathematics of the Universe (WPI), UTIAS, The University of Tokyo, Chiba 277-8583, Japan}

\begin{abstract}
We argue that dark energy with multiple fields is theoretically well-motivated and predicts distinct observational signatures, in particular when cosmic acceleration takes place along a trajectory that is highly non-geodesic in field space. Such models provide novel physics compared to $\Lambda$CDM and quintessence by allowing cosmic acceleration on steep potentials. From the theoretical point of view, these theories can easily satisfy the conjectured swampland constraints and may in certain cases be technically natural, potential problems which are endemic to standard single-field dark energy. Observationally, we argue that while such multi-field models are likely to be largely indistinguishable from the concordance cosmology at the background level, dark energy perturbations can cluster, leading to an enhanced growth of large-scale structure that may be testable as early as the next generation of cosmological surveys.
\end{abstract}

\preprint{YITP-20-111}
\maketitle

\para{Introduction.} Dark energy beyond the cosmological standard model is usually studied in the context of theories with a single scalar field, such as quintessence \cite{Copeland:2006wr} or scalar-tensor gravity \cite{Clifton:2011jh}. While this is primarily motivated by simplicity, physically-realistic models often include additional scalar degrees of freedom, especially if viewed as low-energy effective theories arising from some underlying ultraviolet (UV) completion. For example, compactifications in string theory are characterized by multiple moduli fields, many of which are not necessarily fully stabilized and may therefore play important roles in cosmic evolution at low energies \cite{Douglas:2006es,Arvanitaki:2009fg}. Further theoretical motivation for considering dark energy with multiple fields comes from the recently-proposed \emph{swampland} conjectures~\cite{Obied:2018sgi,Agrawal:2018own,Garg:2018reu,Ooguri:2018wrx}, parameter constraints which, it is claimed,\footnote{It is important to note that the status of these conjectures is unresolved: see, e.g., \rcite{Kachru:2003aw,Akrami:2018ylq,Cicoli:2018kdo,Kachru:2018aqn,Kallosh:2019axr} for counter-arguments.} must be satisfied by any low-energy model which possesses a UV completion in string theory (or sometimes quantum gravity more generally). The swampland bounds on single-field quintessence have been shown to be in strong tension with existing cosmological data \cite{Akrami:2018ylq,Raveri:2018ddi}. These considerations strongly motivate phenomenological attention to multi-field theories, which are common in the inflationary literature,\footnote{The analogy with inflation will prove instructive throughout, particularly because multi-field dynamics is much better studied in the inflationary context \cite{Sasaki:1995aw,GrootNibbelink:2001qt,Lalak:2007vi,Ashoorioon:2009wa,Achucarro:2010jv,Achucarro:2010da,Achucarro:2012sm,Achucarro:2012yr,Pi:2012gf,Achucarro:2017ing,Achucarro:2018vey} than in dark energy, although see, e.g., \rcite{Boyle:2001du,Kim:2005ne,vandeBruck:2009gp,Jimenez:2012iu,Vardanyan:2015oha,Leithes:2016xyh,Akrami:2017cir} for notable examples of multi-field dark energy.} as theoretically-compelling dark energy candidates.

We will argue that there is novel and interesting physics in multi-field dark energy models which follow non-geodesic or \emph{curved} trajectories in field space. As is well-known in the context of inflation, such ``turning" trajectories make accelerated expansion possible in regions where the potential is too steep to otherwise support accelerated expansion \cite{Brown:2017osf,Achucarro:2018vey}. This is in contrast to standard single-field dynamics, which trivially follows a geodesic in the one-dimensional ``field space," and hence has the usual slow-roll requirements. Allowing for this type of strongly multi-field behavior severs the link between a flat potential and an equation of state near $-1$. In addition to opening up an avenue to evade the (conjectured) swampland constraints (see also \rcite{Cicoli:2020cfj,Cicoli:2020noz}), many of which place lower bounds on the slope of the potential, non-geodesic multi-field behavior is a novel physical mechanism for dark energy that leads to observable signatures, predominantly by suppressing the sound speed of fluctuations.

In this Letter, we propose curved trajectories in multi-field theories as a framework for building novel, theoretically well-motivated dark energy models with distinct phenomenological consequences. As a concrete (though non-exhaustive) example, we focus on ``spinning" models, in which the scalars rotate in field space with a nearly-constant speed. While the resultant cosmological background evolution is practically indistinguishable from $\Lambda$CDM (that is, the dark energy equation of state is very close to $-1$), observable features in the evolution of large-scale structure have the potential to distinguish multi-field spinning dark energy from both $\Lambda$CDM and single-field (or single-field-like) quintessence.

From the observational perspective, these models are an essential part of theory space, viewed in the context of interpreting existing and future cosmological data. For single-field models of dark energy, as well as multi-field models with shallow potentials, observing a constant dark energy equation of state $w_\mathrm{DE}$ would require the fields to be effectively non-dynamical. As we show in this Letter, for multi-field models with non-geodesic trajectories it is possible to have $w_\mathrm{DE}$ arbitrarily close to $-1$ even if the fields are highly dynamical. This means that a $\Lambda$-like equation of state, if supported by the next generation of cosmological surveys, would not necessarily imply that the late-time cosmic acceleration is driven by a non-dynamical cosmological constant.

The novel physical effects of multi-field dark energy are more pronounced at the level of perturbations. In order for a single field to drive cosmic acceleration, the relevant mass scale typically must be of order $H_0$, the present-day expansion rate. The associated Compton wavelength is therefore around the size of the horizon, preventing dark energy from clustering on observable (sub-horizon) scales. We will show that the models we consider here are a type of \emph{clustering dark energy} \cite{Bean:2003fb,Weller:2003hw,Mota:2004pa,Nunes:2004wn,Hu:2004yd,Hu:2004kh,Takada:2006xs,Creminelli:2008wc,Creminelli:2009mu,Amendola:2016saw,Heneka:2017ffk,Hassani:2020buk}: the sound speed of dark energy fluctuations in these models is much smaller than unity for a wide range of parameters, so the sound horizon can be much smaller than the cosmological horizon, leading to clustering at observable (sub-horizon) scales.\footnote{By contrast, in order for a single field to drive cosmic acceleration, the relevant mass scale typically must be of order $H_0$, the present-day expansion rate. The associated Compton wavelength is therefore around the size of the horizon, preventing single-field dark energy from clustering at smaller scales.} We therefore expect significant enhancements in clustering of large-scale structure at low redshifts. This feature provides a powerful method of testing this important class of dark energy models against a cosmological constant, as well as more orthodox, slowly-rolling dark energy models.

\para{Multi-field dark energy.} We are interested in dark energy models with multiple scalar fields minimally coupled to gravity. At leading order in derivatives, we consider a standard $\sigma$-model setup,
\begin{align}
\mathcal{L} = \frac{\Mp^2}{2}R -\frac{1}{2}\mathcal{G}_{ab}(\phi)\partial_\mu \phi^a \partial^\mu \phi^b - V(\phi) + \mathcal{L}_\mathrm{m}\,,
\end{align}
where $\mathcal{G}_{ab}$ is the field-space metric, which is allowed to depend on the fields $\phi^a$, $V(\phi)$ is the potential, and $\mathcal{L}_\mathrm{m}$ is the matter Lagrangian.

Restricting ourselves to cosmological solutions with a Friedmann-Lema\^itre-Robertson-Walker (FLRW) metric, and adopting the framework developed in \rcite{Sasaki:1995aw,GrootNibbelink:2001qt,Achucarro:2010jv}, the scalar field equations of motion are
\begin{align}\label{eq:eoms}
    D_t\dot\phi^a + 3H\dot\phi^a + V^a = 0\,,
\end{align}
where $V_a \equiv \partial V/\partial \phi^a$, $H\equiv \dot a/a$ is the Hubble rate, overdots denote derivatives with respect to cosmic time $t$, and the field-space covariant time derivative $D_t$ is defined by
\begin{align}\label{eq:cov_der}
D_tA^a \equiv \dot A^a + \Gamma^a_{bc}A^b\dot\phi^c\,,
\end{align}
with $\Gamma^a_{bc}$ the field-space Christoffel symbols. The Friedmann equation is
\begin{align}\label{eq:Friedmann}
    3\Mp^2H^2 = \frac{1}{2}\dot\phi^2 + V + \rho_\mathrm{M} + \rho_\mathrm{R}\,,
\end{align}
where we have defined
\begin{align}
    \dot\phi^2 \equiv \mathcal{G}_{ab}\dot\phi^a\dot\phi^b\,,
\end{align}
which characterizes the speed along the background trajectory in field space, and denoted the matter and radiation energy densities by $\rho_\mathrm{M}$ and $\rho_\mathrm{R}$, respectively.

We will find it convenient to introduce the normalized tangent and normal vectors to the field-space trajectory,
\begin{align}
\mathcal{T}^a &\equiv \frac{\dot\phi^a}{\dot\phi}\,,\\
\mathcal{N}^a &\equiv -\frac{1}{\Omega}D_t\mathcal{T}^a\,,
\end{align}
with
\begin{equation}
\Omega \equiv |D_t\mathcal{T}|.
\end{equation}
Projecting the scalar equation of motion \eqref{eq:eoms} along these directions, we find
\begin{align}\label{eq:eom_tangent}
    \ddot\phi + 3H\dot\phi + V_\mathcal{T} &= 0\,,\\
    V_\mathcal{N} &= \dot\phi\Omega\,,\label{eq:eom_normal}
\end{align}
where we have defined
\begin{equation}
V_\mathcal{T} \equiv V_a\mathcal{T}^a,\qquad V_\mathcal{N} \equiv V_a\mathcal{N}^a.
\end{equation}

The novelty of multi-field dark energy hinges on the fact that the fields need not follow geodesic trajectories in field space. The degree of departure from a geodesic trajectory, or \emph{turning}, is characterized by $\Omega\equiv|D_t\mathcal{T}|$, as the geodesic equation is $D_t\mathcal{T}^a=0$. In order to compute $\Omega$ along cosmological trajectories, we will need to express it in terms of the fields $\phi^a$. For two-field systems, as we will consider in this Letter, we can write the normal as $\mathcal{N}^a=\varepsilon^{ab}\mathcal{T}_b$, with $\varepsilon_{12}=\sqrt{\det\mathcal{G}}$. It follows from \cref{eq:eom_normal} and the definition of $V_\mathcal{N}$ that
\begin{align}
    \Omega^2 = \frac{1}{\det\mathcal{G}}\frac{\left(\dot\phi_1V_2 - \dot\phi_2V_1\right)^2}{\dot{\phi}^4}\,, \label{eq:Omega2D}
\end{align}
where $\dot\phi_a=\mathcal{G}_{ab}\dot\phi^b$. For an arbitrary number of scalars, this generalizes to
\begin{align}
    \Omega^2 = \frac{\mathcal{F}_{ab}\mathcal{F}^{ab}}{2\dot{\phi}^4}\,,
\end{align}
where
\begin{equation}
\mathcal{F}_{ab} \equiv \dot\phi_aV_b - \dot\phi_bV_a.
\end{equation}
To see this, note that $\mathcal{F}_{ab}=2\dot\phi \mathcal{T}_{[a}V_{b]}=2V_\mathcal{N}\mathcal{T}_{[a}\mathcal{N}_{b]}$, where we have decomposed $V_a=V_\mathcal{T}\mathcal{T}_a+V_\mathcal{N}\mathcal{N}_a$.\footnote{That $V_a$ can be decomposed this way is obvious for two fields. In general it follows from \cref{eq:eoms}.} Squaring $\mathcal{F}_{ab}$ and using \cref{eq:eom_normal}, the result follows.

To explain our mechanism, it is convenient to first ignore matter and focus on dark energy domination, in which case the physical picture is similar to multi-field inflation. Cosmic acceleration occurs when $H$ is nearly constant, i.e., $\epsilon\ll1$ with $\epsilon\equiv-\dot H/H^2$ the \emph{Hubble slow-roll parameter}. While a single canonical scalar requires a flat potential in order to drive a period of acceleration, in the presence of multiple fields there can also be acceleration due to large \emph{turning}, even in regions where the potential is steep. Defining the \emph{potential slow-roll parameter} $\epsilon_V$ as
\begin{align}
    \epsilon_V \equiv \frac{\Mp^2}{2}\frac{\mathcal{G}^{ab}V_aV_b}{V^2} = \frac{\Mp^2}{2}\frac{V_\mathcal{N}^2 + V_\mathcal{T}^2}{V^2}\,,\label{eq:epsilonV}
\end{align}
and considering the slow-roll r\'egime $\epsilon\ll1$ (but \emph{not} necessarily $\epsilon_V\ll1$), it is straightforward to show that \cite{Hetz:2016ics,Achucarro:2018vey}
\begin{align}\label{eq:consistency_relation_inf}
    \epsilon = \epsilon_V\left(1 + \frac{\Omega^2}{9H^2}\right)^{-1}\,.
\end{align}
The essential insight for the models we consider is that, for a sufficiently non-geodesic trajectory, $\Omega\gg H$, the scalars can drive accelerated expansion even when they are in a steep region of the potential.

This phenomenon underlies many of the novel observational signatures of multi-field inflation, as well as its avoidance of the swampland bounds that plague single-field theories (see also \rcite{Solomon:2020viz}). Our aim is to investigate its utility for dark energy model-building. The main difference with the inflationary case is the presence of other matter fields with significant energy densities. Instead of demanding $\epsilon\ll1$ (which will not hold during radiation and matter domination), we want to find the conditions under which the scalars' energy density changes slowly, i.e., $\epsilon_\mathrm{DE}\ll1$, with
\begin{align}
    \epsilon_\mathrm{DE} = \frac{3}{2}(w_\mathrm{DE} + 1)=\frac{3}{2}\frac{\dot\phi^2}{\frac{1}{2}\dot\phi^2+V}\,,
\end{align}
where $w_\mathrm{DE}$ is the dark energy equation of state. When dark energy dominates, $\epsilon_\mathrm{DE}$ approaches $\epsilon$. Repeating the steps that led to \cref{eq:consistency_relation_inf}, we find, assuming $\epsilon_\mathrm{DE}\ll1$,
\begin{align}\label{eq:consistency_relation}
    \epsilon_\mathrm{DE} = \epsilon_V\Omega_\mathrm{DE}\left(1 + \frac{\Omega^2}{9H^2}\right)^{-1}\,,
\end{align}
where $\Omega_\mathrm{DE}\equiv\rho_\phi/\rho_\mathrm{tot}\approx V/(3\Mp^2H^2)$ is the usual dark energy density parameter. We see that the presence of additional matter fields can only suppress $\epsilon_\mathrm{DE}$ further.

We conclude that multi-field dark energy, much like inflation, can drive accelerated expansion on arbitrarily steep potentials as long as $\phi^a$ follows a highly non-geodesic path in field space, such as a spinning trajectory. Severing the link between the slope of the potential and cosmic acceleration allows these theories to potentially evade problems endemic to single-field theories. In addition to the aforementioned swampland conjectures, the flatness of the potential is typically controlled by a small parameter which is not stable against radiative corrections \cite{Kolda:1998wq}, a problem made particularly acute if the swampland bounds are imposed \cite{Hertzberg:2018suv}. Observationally, we will see that, while multi-field dark energy is practically indistinguishable from a cosmological constant at the background level, it can provide a distinct, rich, and novel phenomenology for structure formation.

\para{A concrete example.} In order to illustrate the multi-field dark energy mechanism as simply as possible, we restrict ourselves to two fields with a polar parametrization, $\phi^a=(r,\theta)$, and impose $U(1)$ invariance through the shift symmetry $\theta\to\theta+c$. The most general U(1)-invariant field-space metric is
\begin{align}\label{eq:model_metric}
    \mathcal{G}_{ab} = \mathrm{diag} (1, f(r)),
\end{align}
with Ricci curvature
\begin{equation}
	\mathcal{R} = -\frac{1}{f}\left(f'' - \frac12\frac{f'^2}{f}\right),
\end{equation}
where primes denote $r$ derivatives. We will leave $f(r)$ general, though for numerical illustrations we will choose a flat field space, $f(r)=r^2$.\footnote{This choice is used in ``spintessence" \cite{Boyle:2001du}, where the two fields form a complex scalar $\Phi=re^{i\theta}$ with a canonical kinetic term $|\partial\Phi|^2$.}

In the potential, $U(1)$ invariance ($V=V(r)$) turns out to be incompatible with our proposed mechanism and must therefore be broken. To see this, note that the symmetry implies a conserved charge,
\begin{equation}
Q=a^3f(r)\dot\theta. \label{eq:cons-charge}
\end{equation}
In flat space, $a=1$ and there exist stable circular orbits, but in an expanding universe this is not possible: circular orbits decay as $\dot\theta\sim a^{-3}$ goes to zero and $r$ falls to the minimum of its potential.\footnote{This holds for an arbitrary number $N$ of scalars when $\mathcal{G}_{ab}$ and $V$ depend on a single field: there are $N-1$ conserved charges, each decaying as $a^{-3}$.} To avoid this problem, the potential must depend on $\theta$. While we keep $V(r,\theta)$ general when possible, when we need a concrete model we will borrow from the inflationary literature \cite{Achucarro:2012yr} a potential which breaks $U(1)$ as softly as possible,
\begin{align}\label{eq:model_pot}
    V(r, \theta) = V_0 - \alpha\theta + \frac{1}{2}m^2(r - r_0)^2\,,
\end{align}
with $V_0$, $\alpha$, $m$, and $r_0$ free parameters.

As an aside, we note that $\alpha$ is radiatively stable, as $U(1)$ invariance is restored in the $\alpha\to0$ limit. If the field space is flat and $r_0=0$, then the effective $r$ mass $m_r^2\equiv m^2+(\partial\theta)^2$ on a particular background may also be protected from large quantum corrections. While the ``old" cosmological constant problem, i.e., the radiative instability of $V_0$, remains, and requires additional physics to address \cite{Weinberg:1988cp,Dvali:2007kt,Khoury:2018vdv}, as is typically the case even for technically-natural dark energy theories (e.g., \cite{DAmico:2018mnx}), the mechanism presented here provides a promising route towards the construction of dark energy with enhanced naturalness properties.

On a cosmological background, the equations of motion are\footnote{For a dynamical-systems analysis of these equations for restricted choices of $V=V(r)$, see \rcite{Cicoli:2020cfj,Cicoli:2020noz}.}
\begin{align}
3\Mp^2H^2 = \frac{1}{2}\left( \dot{r}^2 + f\dot{\theta}^2 \right) + V + \rho_\mathrm{M} + \rho_\mathrm{R}\,,\label{eq:background1}\\
\ddot{r} + 3H\dot{r} + V_r - \frac{1}{2}f'\dot{\theta}^2 = 0\,,\label{eq:background2}\\
\ddot{\theta} + 3H\dot{\theta} + \frac1fV_\theta + \frac{f'}{f}\dot{r}\dot{\theta} = 0\,,\label{eq:background3}
\end{align}
where primes denote $r$ derivatives. The $r$ equation of motion \eqref{eq:background2} has the usual forcing term $V_r$, which pulls $r$ down towards the minimum of its potential (as in standard quintessence), as well as a $-\frac12f'\dot{\theta}^2$ term which drives $r$ up the potential. Our mechanism relies on balancing these competing forces by having the fields spin with $\dot\theta^2\approx2V_r/f'$. For the field-space metric \eqref{eq:model_metric} with $f=r^2$ and potential \eqref{eq:model_pot}, this amounts to a solution with $r$ approximately constant and
\begin{equation}
\dot\theta^2\approx m^2\left(1-\frac{r_0}{r}\right).\label{eq:thetadot}
\end{equation}

It is easy to show that the combination of \cref{eq:Omega2D,eq:thetadot} implies $\Omega^2=\dot\theta^2$ on this trajectory. In the inflationary context, such circularly spinning solutions are cosmological attractors \cite{Achucarro:2012yr}. In order to check whether this mechanism is also viable for dark energy, we include matter and radiation and solve for the resultant cosmologies numerically.\footnote{The codes used in this paper are publicly available at \url{https://github.com/valerivardanyan/Multifield-Dark-Energy}.} We generically find solutions which realize the proposed mechanism: despite being significantly displaced from the minimum of its potential at $r=r_0$, cosmic history is quantitatively very close to $\Lambda$CDM.

\begin{figure}[hpt!]
    \centering
    \includegraphics[width=0.5\textwidth]{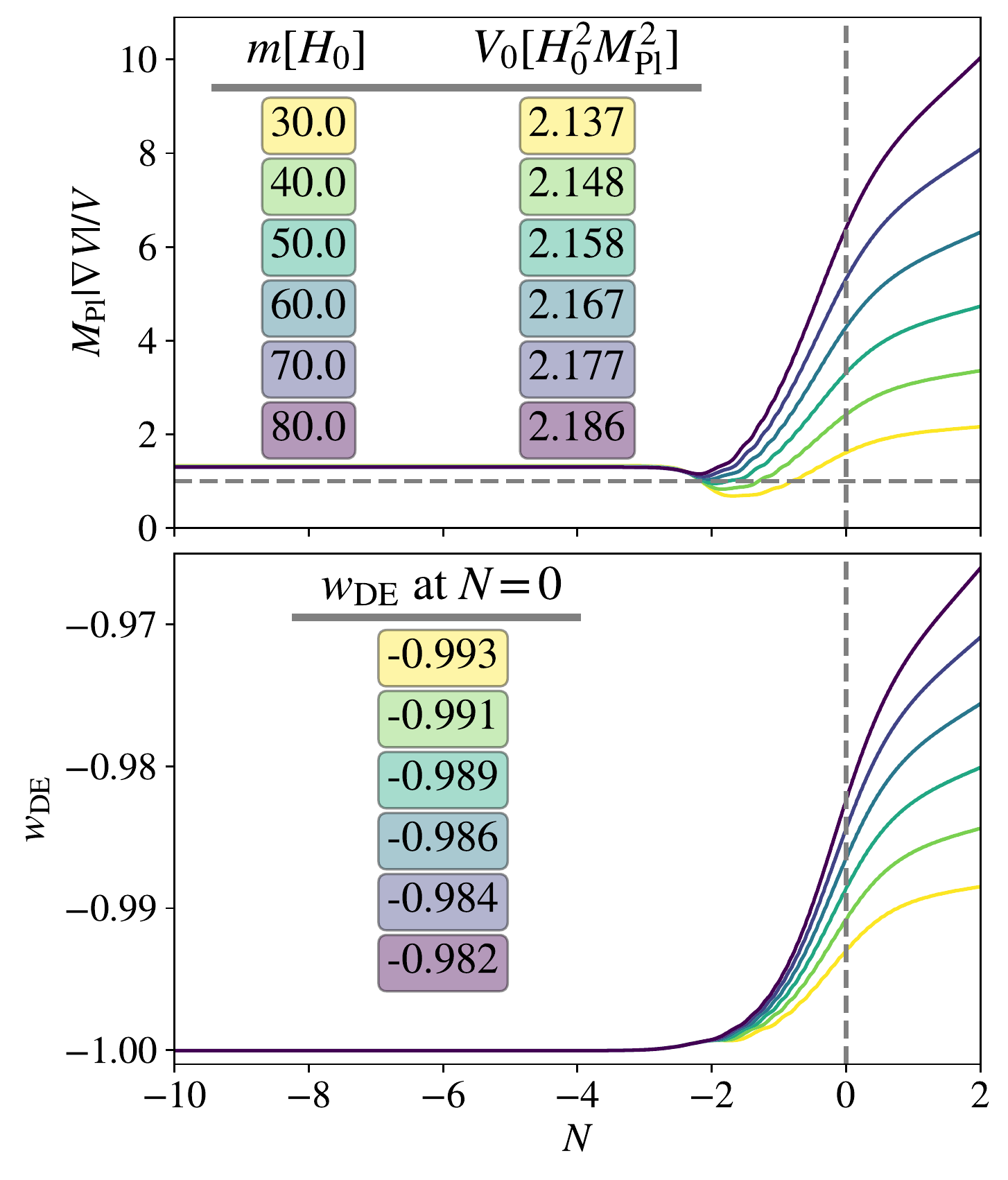}
    \caption{Time evolution of the slope of the potential $\Mp |\nabla V|/V$ along the trajectory (upper panel) and of the dark energy equation of state $w_\mathrm{DE}$ (lower panel) for the two-field model with a flat field-space metric and the potential \eqref{eq:model_pot}. We have chosen $\alpha/H_0^2M_\mathrm{Pl}^2 = 2\times 10^{-3}$, $r_0/M_\mathrm{Pl} = 7\times 10^{-4}$ and have varied $m$ (specified in the upper panel). For each choice of $m$ we have picked $V_0$ such that the spatial curvature of the universe vanishes. There is no strong dependence on initial conditions. Our time variable is the number of $e$-foldings $N \equiv \ln(a)$, with $N=0$ corresponding to the present.}
    \label{fig:eos_with_matter}
\end{figure}

This behavior is demonstrated, for representative parameters, in \cref{fig:eos_with_matter}. In the upper panel we plot the gradient of the potential over time, showing that, as promised, the dark energy lives in a steep region of the potential, while in the lower panel we plot the evolution of $w_\mathrm{DE}$ over cosmic history, finding that it is extremely close to $-1$. The combination of these is a unique signature of multi-field dark energy. We have also checked that the swampland condition $\Mp \vert\nabla V\vert/V\gtrsim\mathcal{O}(1)$ is satisfied over the entirety of field space,\footnote{Strictly speaking $\epsilon_V$ eventually becomes small at very large $r$, but the potential \eqref{eq:model_pot} should be properly viewed as part of an effective field theory and so cannot be trusted at arbitrarily large field values.} as required by the swampland conjectures (which are more restrictive than just being true along the cosmological trajectory).

While we expect $w_\mathrm{DE}\approx-1$ during dark energy domination, as the physics is similar to multi-field inflation, we see from \cref{fig:eos_with_matter} that this equation of state also holds during the matter- and radiation-dominated eras. During these epochs, Hubble friction dominates the forcing terms in \cref{eq:background2,eq:background3}, causing $r$ and $\theta$ to freeze. As matter and radiation dilute away, the Hubble friction becomes smaller than the forcing terms and the fields start to roll. The $r$ field falls slightly down the potential before stabilizing as $\theta$ spins up, transitioning into the spinning r\'egime and the onset of dark energy domination. During this period, the dark energy equation of state starts evolving from its frozen value of $-1$, but only slightly: in the spinning r\'egime, the Hubble slow-roll parameter $\epsilon_\mathrm{DE}$ is suppressed by the turning rate, cf. \cref{eq:consistency_relation}, so that $w_\mathrm{DE}$ remains close to $-1$.

Observationally, this model is to some extent a victim of its own success: it mimics $\Lambda$CDM so efficiently that it is unlikely to be distinguishable from the concordance cosmology at the background level, even with qualitatively rather different physics than $\Lambda$CDM or standard slow-roll quintessence. Forecast analyses from the forthcoming Stage IV cosmological surveys predict percent-level constraints on parameters like the dark energy equation of state. The \textit{Euclid} space mission~\cite{Laureijs:2011gra}, an important representative of these surveys, is expected to measure the present value of $w_\mathrm{DE}$, commonly denoted as $w_0$, with at best a $1\sigma$ uncertainty of $\sigma_{w_0} \approx 0.025$~\cite{Blanchard:2019oqi}. As seen in the lower panel of \cref{fig:eos_with_matter}, \textit{Euclid} will not be able to distinguish $w_\mathrm{DE}$ in the dark energy model proposed here from the $\Lambda$CDM value of $-1$, although in principle there may be regions of parameter space in which $w_\mathrm{DE}+1$ is just large enough to be observable.

We emphasize that the model discussed in this section serves as a minimal working example. Our results do not depend strongly on the details of the model, and we expect them to be qualitatively robust for any potential that supports strongly non-geodesic motion in field space.\footnote{Minor quantitative details, such as the behavior of the fields when they are subdominant, can change from model to model. For instance, by appropriately changing the $\theta$ potential, the system may enter a scaling r\'egime during the matter-dominated era, in analogy with the single-field model in \rcite{Copeland:1997et}. This allows the transition from frozen to spinning behavior to occur more quickly, as the $\theta$ field is already dynamical when dark energy becomes dominant. While illustrative, this scenario is somewhat contrived, and the $\theta$ shift symmetry would no longer be broken softly.}
  
\para{Clustering dark energy.} While the dark energy mechanism proposed here is likely to be observationally indistinguishable from $\Lambda$CDM at the background level, the story changes dramatically when we consider perturbations. In this section we briefly discuss why one should expect these theories to produce novel signatures in structure formation, while saving a full analysis of perturbations and the comparison to observations for future work.

For a wide range of parameters, the sound speed of fluctuations is heavily suppressed, leading to \emph{clustering} dark energy. To see this, we expand the fields around their background values as $\phi^a=\bar\phi^a+\delta\phi^a$. It is convenient to work with the field fluctuations parallel and perpendicular to the background trajectory,
\begin{equation}
\delta\phi^\mathcal{T}\equiv\mathcal{T}_a\delta\phi^a,\qquad \delta\phi^\mathcal{N}\equiv\mathcal{N}_a\delta\phi^a.
\end{equation}
Working with the Newtonian gauge for scalar metric perturbations,
\begin{equation}
g_{\mu\nu} = \mathrm{diag}\left(-a^2(1 - 2\Phi), \delta_{ij}a^2(1 + 2\Phi)\right),
\end{equation}
where $\Phi$ is the gravitational potential, and including dark matter fluctuations $\delta_\mathrm{DM}$, the full linearized Einstein equations become
\begin{widetext}
\begin{align}
	&\delta\phi^{\mathcal{T}\prime\prime} + 2\mathcal{H}\delta\phi^{\mathcal{T}\prime} + 2a\Omega\delta\phi^{\mathcal{N}\prime} + \left[k^2 + a^2\mathcal{D}^2_\mathcal{T} V \right]\delta\phi^\mathcal{T} + 4a\mathcal{H}\Omega\delta\phi^\mathcal{N}  - 2a^2\Phi \mathcal{D}_\mathcal{T} V+ 4\Phi^\prime\phi^{\prime}=0,\label{eq:deltaphiT_full}\\
	&\delta\phi^{\mathcal{N}\prime\prime} + 2\mathcal{H}\delta\phi^{\mathcal{N}\prime} - 2a\Omega\delta\phi^{\mathcal{T}\prime} + \left[k^2 + \Meff^2\right]\delta\phi^\mathcal{N} - 2a\Phi \phi^\prime\Omega=0,\label{eq:deltaphiN_full}\\
         &\Phi^{\prime\prime} + 6\mathcal{H}\Phi^{\prime} + \left(4\mathcal{H}^2 + 2\mathcal{H}^\prime\right)\Phi + k^2\Phi=\frac{a^2}{2M^2_\mathrm{Pl}}\left(\rho_\mathrm{DM}\delta_\mathrm{DM} + 2V_\mathcal{T}\delta\phi^\mathcal{T} + 2V_\mathcal{N}\delta\phi^\mathcal{N}\right),
\end{align}
\end{widetext}
where primes denote conformal time derivatives, $\mathcal{H}$ is the conformal time Hubble rate, $\rho_\mathrm{DM}$ is the background dark matter density, $V_\mathcal{T} \equiv \mathcal{D}_a\mathcal{T}^a V$ and $V_\mathcal{N} \equiv \mathcal{D}_a\mathcal{N}^aV$, where $\mathcal{D}_a$ is the covariant derivative associated to $\mathcal{G}_{ab}$, and the effective $\delta\phi^\mathcal{N}$ mass is
\begin{align}
\Meff^2 \equiv a^2V_\mathcal{NN} - a^2\Omega^2 + \mathcal{R}\frac{\phi^{\prime 2}}{2}. \label{eq:Meff}
\end{align}
Here $V_\mathcal{NN} \equiv \mathcal{N}^a\mathcal{N}^b\mathcal{D}_a\mathcal{D}_b V$ and $\mathcal{R}$ is the Ricci scalar for $\mathcal{G}_{ab}$. For concreteness, we focus on scalar field perturbations on scales smaller than the sound horizon and ignore dark matter fluctuations and gravitational backreaction, which suffices to illustrate the important physical effects. In this limit, the scalar equations of motion \cref{eq:deltaphiT_full} and \cref{eq:deltaphiN_full} are
\begin{align}
	&\delta\phi^{\mathcal{T}\prime\prime} + k^2\delta\phi^\mathcal{T} = - 2a\Omega\delta\phi^{\mathcal{N}\prime}, \label{eq:deltaphiT}\\
	&\delta\phi^{\mathcal{N}\prime\prime} + \left(k^2 + \Meff^2\right)\delta\phi^\mathcal{N}  = 2a\Omega\delta\phi^{\mathcal{T}\prime}.\label{eq:deltaphiN}
\end{align}
Note that we have neglected a small mass term in \cref{eq:deltaphiT}, which is necessarily suppressed in the spinning r\'egime where $\ddot\phi\ll H\dot\phi$.

We see from \cref{eq:deltaphiT,eq:deltaphiN} that a non-zero $\Omega$ introduces a coupling between $\delta\phi^\mathcal{T}$ and $\delta\phi^\mathcal{N}$. To identify the propagating degrees of freedom, we look for solutions with time dependence $e^{i\omega\tau}$ to obtain the dispersion relation,
\begin{equation}\label{eq:dispersion}
	\omega^4 - \omega^2\left(2k^2 + \Meff^2 + 4a^2\Omega^2\right) + k^2\left(k^2 + \Meff^2\right) = 0.
\end{equation}
On geodesic trajectories, $\Omega=0$ and the dispersion relation factorizes, $(\omega^2-k^2)(\omega^2-k^2-\Meff^2)=0$, from which we can identify a light mode and a heavy mode of mass $\Meff$, each propagating at the speed of light. Including $\Omega$, the full dispersion relations are
\begin{align}
\omega_\pm^2&=\frac{\Meff^2+4a^2\Omega^2}{2}+k^2\nonumber\\
&\hphantom{{}=}\pm\sqrt{\left(\frac{\Meff^2+4a^2\Omega^2}{2}\right)^2+4a^2\Omega^2k^2}.
\end{align}
The light mode corresponds to $\omega_-$, since $\omega_-\to0$ as $k\to0$. Considering scales larger than the Compton wavelength of the heavy mode, $k^2\ll\Meff^2+4a^2\Omega^2$, and expanding the light-mode dispersion relation to leading order, we see that it propagates with a modified sound speed,
\begin{equation}
\omega_-^2 = c_\mathrm{s}^2k^2+\mathcal{O}(k^4),
\end{equation}
where
\begin{equation}
c_\mathrm{s}^{-2} \equiv 1+\frac{4a^2\Omega^2}{\Meff^2}.
\end{equation}
The sound speed is suppressed when $a^2\Omega^2\gg\Meff^2$, which per \cref{eq:Meff} requires $\Omega^2\approx V_\mathcal{NN}+\mathcal{R}\dot\phi^2/2$. To illustrate quantitatively the typical scales involved, we take as an example $m=30H_0$ (cf. \cref{fig:eos_with_matter}), for which we have, in the present day, $a^2\Omega^2/H_0^2 \approx 750$ and $M_\mathrm{eff}^2/H_0^2 \approx 150$, with $c_\mathrm{s}^{2}\approx 0.047$.

While this r\'egime may seem highly tuned, it is in fact supported in the model discussed above, as we have confirmed numerically for the parameter choice in \cref{fig:eos_with_matter}. This $c_\mathrm{s}^2\ll1$ attractor is well-known in the inflationary context, and exists as long as the potential parameters satisfy \cite{Achucarro:2012yr}
\begin{equation}
1\gg r_0 V_0^{1/4}\sqrt{\frac{m}{\alpha \Mp}}\gg\frac{V_0}{m^2\Mp^2}\gg\frac{\alpha}{m \Mp \sqrt{V_0}}.
\end{equation}
During dark energy domination, the same equations of motion apply and so the same attractor is present.

For completeness we present the heavy mode's dispersion relation to quadratic order in $k$,
\begin{align}
\omega_+^2 &= \Meff^2+4a^2\Omega^2+\frac{\Meff^2+8a^2\Omega^2}{\Meff^2+4a^2\Omega^2}k^2+\mathcal{O}(k^4)\nonumber\\
&= \frac{\Meff^2}{c_\mathrm{s}^2} + (2-c_\mathrm{s}^2)k^2 + \mathcal{O}(k^4).
\end{align}
Recall that $\Meff$ is this mode's mass in the geodesic limit $\Omega\to0$. Spinning increases the mass by a factor of $c_\mathrm{s}^{-2}$. This suggests a wide range of intermediate scales, $H^2\ll k^2/a^2\ll\Meff^2c_\mathrm{s}^{-2}$, where the heavy mode can be integrated out, leading to a simpler single-field effective theory, as in inflation \cite{Achucarro:2012sm,Achucarro:2012yr}, while remaining in the sub-horizon r\'egime. We note speculatively that a condensate of this heavy field could potentially be a dark matter candidate; we leave a more detailed analysis of this possibility to future work.

In regions of parameter space where the sound speed is suppressed, we expect enhanced structure formation in the late universe, as there is a well-known correspondence between a small sound speed and dark energy clustering \cite{Hu:2004yd,Takada:2006xs,Creminelli:2008wc,Creminelli:2009mu}. The physical explanation is that a reduced speed of sound pushes the Jeans instability to sub-horizon scales. Modes of the light field with larger wavelengths will therefore cluster on observationally accessible scales.\footnote{While \cref{eq:deltaphiT,eq:deltaphiN} hold only below the sound horizon, the sound speed remains small at all scales $k^2/a^2 \ll \Meff^2 c_\mathrm{s}^{-2}$, where the heavy mode can be integrated out \cite{Achucarro:2012yr}.} This is in contrast to canonical, single-field dark energy, where the Jeans scale is super-horizon, so the Jeans instability is not observable.

In brief we mention two other reasons to expect clustering or other interesting features in structure formation in theories of the type discussed here, depending on the details of the model:
\begin{enumerate}
\item While in the inflationary context the heavy mode is suppressed in amplitude \cite{Achucarro:2012yr}, it could in principle (depending on initial conditions) be non-negligible in the late universe. This situation would be similar to a canonical massive scalar field, which clusters on sub-horizon scales when its mass is larger than the Hubble scale (cf., e.g., \rcite{Nambu:1989kh,Amendola:2015ksp}).
\item The field space curvature $\mathcal{R}$ contributes to $\Meff$, and a sufficiently negative curvature can render the heavy mode tachyonic. During inflation this phenomenon is known as geometrical destabilization and is considered problematic, spoiling otherwise successful models \cite{Renaux-Petel:2015mga}. In context of the late universe, however, a mild tachyonic instability might imprint unique features on the dark matter distribution, opening up a new window for probing the curvature of field space.
\end{enumerate}

\para{Conclusions.} In this Letter we have proposed a novel class of multi-field dark energy models where the fields do not follow geodesic trajectories in field space, allowing steep potentials to lead to cosmic acceleration. We have argued why these models are theoretically well-motivated when new developments in high energy physics and quantum gravity are considered. By focusing on a concrete and representative example, we have studied the cosmological background evolution in these models and shown that they are practically indistinguishable from the standard $\Lambda$CDM model. This means that constraining the equation of state of dark energy by next-generation cosmological surveys to values arbitrarily close to $-1$ will not exclude the possibility of dark energy being highly dynamical. We have argued, however, that our models do result in features in formation and evolution of the cosmic large-scale structure, which can potentially distinguish the models from $\Lambda$CDM and single-field dark energy. We have derived and presented the full linearized Einstein equations where gravitational effects and dark matter fluctuations are included. By restricting ourselves to subhorizon scales and only considering dynamics of the scalar field fluctuations in the absence of gravitational backreaction, we have demonstrated that in theories of the type discussed in this Letter dark energy is expected to cluster during cosmic structure formation. We have identified three reasons for this clustering: sound speed suppression for the light mode of dark energy fluctuations, non-negligibility of the heavy mode, and tachyonic instability of the heavy mode resulting from a negative field space curvature. Each of these effects on structure formation requires a more detailed numerical analysis to derive quantitative predictions for cosmological observables, in particular in the context of clustering dark energy~\cite{Bean:2003fb,Weller:2003hw,Mota:2004pa,Nunes:2004wn,Hu:2004yd,Hu:2004kh,Takada:2006xs,Creminelli:2008wc,Creminelli:2009mu,Amendola:2016saw,Heneka:2017ffk,Hassani:2020buk}. This is beyond the scope of this Letter and we leave it for future work.

\begin{acknowledgments}
We thank Martin Bucher, Edmund J. Copeland, Nick Kaiser, S\'ebastien Renaux-Petel, and Benjamin D. Wandelt for helpful discussions. Y.A. is supported by LabEx ENS-ICFP: ANR-10-LABX-0010/ANR-10-IDEX-0001-02 PSL*. M.S. is supported in part by JSPS KAKENHI No. 20H04727. A.R.S. is supported by DOE HEP grants DOE DE-FG02-04ER41338 and FG02-06ER41449 and by the McWilliams Center for Cosmology, Carnegie Mellon University. V.V. is supported by the WPI Research Center Initiative, MEXT, Japan.
\end{acknowledgments}

\bibliography{refs}

\end{document}